\documentclass{PoS}

\usepackage{graphicx}% Include figure files
\usepackage{dcolumn}% Align table columns on decimal point
\usepackage{bm}% bold math
\usepackage{amssymb}
\usepackage{amsmath}
\usepackage{epsfig}
\newcommand\Tr{{\rm Tr\,}}

\newcommand\beq{\begin{eqnarray}}
\newcommand\eeq{\end{eqnarray}}

\newcommand\Eq[1]{Eq. (\ref{eq:#1})}

\newcommand\Fig[1]{Fig. (\ref{fig:#1})}
\newcommand\Tab[1]{Table (\ref{tab:#1})}

\newcommand{\bfx}{{\bf x}}

\newcommand{\calN}{{\cal N}}

\title{%
Numerical simulation of $\calN=1$ supersymmetric Yang-Mills theory
}

\ShortTitle{%
Numerical simulation of $\calN=1$ supersymmetric Yang-Mills theory
}

\author{%
\speaker{Michael G. Endres} \\
Columbia University \\
E-mail: \email{mge2112@columbia.edu}
}

\abstract{%
We present results from a lattice study of $SU(2)$ color, $\calN=1$ supersymmetric Yang-Mills theory using domain wall fermions.
Supersymmetry in this particular lattice formulation is expected to emerge in the continuum and chiral limits without any fine-tuning of operators.
Preliminary results for the static quark potential, residual mass, chiral condensate and spectrum--a potential indicator of supersymmetry restoration--are presented and discussed. 
}

\FullConference{%
The XXVI International Symposium on Lattice Field Theory \\
July 14 - 19, 2008\\
Williamsburg, Virginia, USA
}

\begin{document}

\section{Introduction}
In recent years, much effort has been devoted toward formulating supersymmetric (SUSY) gauge theories on the lattice.
This has been partially motivated by the theoretical as well as technical challenges associated with the problem, as well as the obvious potential role of SUSY in beyond the standard model physics.
Of crucial importance pertaining to the latter point is an understanding of dynamical symmetry breaking--something which may in principle be achieved with numerical simulations provided an appropriate lattice discretization of the theory is found.

Since na\'ive discretizations typically break SUSY explicitly, lattice simulations require fine-tuning in order to cancel off the undesirable SUSY breaking operators which may arise through radiative corrections.
However, it has been realized for some time that one of the simplest SUSY theories, $\calN=1$ supersymmetric Yang-Mills (SYM), may be simulated with conventional lattice discretizations and yet require only a minimal degree of fine-tuning.
Although this theory does not possess SUSY breaking (as implied by a nonvanishing Witten index \cite{Witten:1982df}), it is believed to exhibit a variety of other interesting features such as discrete chiral symmetry breaking.
The field content of $\calN=1$ SYM consists of a vector field and a single adjoint representation Majorana fermion, and in conventional lattice discretizations of this theory, the only relevant SUSY violating operator which may arise radiatively is a gluino mass term.
As a result, in the chiral and continuum limits SUSY is restored {\it accidentally} at infinite space-time volume.

In the past, a variety of numerical studies have employed Wilson fermions in order to simulate $\calN=1$ SYM (for a review, see \cite{Feo:2002yi}), however, these were subject to both difficulties of fine-tuning and the sign problem.
In contrast, it was observed in \cite{Kaplan:1999jn} that domain-wall fermions (DWFs) are an ideal fermion discretization for simulating $\calN=1$ SYM because of their good chiral properties \cite{Kaplan:1992bt,Narayanan:1992wx,Shamir:1993zy} and positivity of the fermion Pfaffian obtained from ``integrating out'' the fermion degrees of freedom.
In this formulation, the chiral limit may be achieved without any fine-tuning of operators.

The first and until recently the only study to use DWFs to investigate $\calN=1$ SYM focused on the chiral limit of the gluino condensate \cite{Fleming:2000fa}.
In our work, we expand on this early study in several important respects: we
1) establish the lattice scale by measuring the static quark potential and provide evidence for confinement which is consistent with expectations,
2) determine the size of the residual mass in order to ascertain the proximity to the SUSY point,
3) extrapolate the chiral condensate to the chiral limit using a recent, theoretically motivated fit formula for its $L_s$ dependence, and
4) study the spectrum of the theory. 
With exception to the third point, questions such as these could not be easily addressed in \cite{Fleming:2000fa} because space-time volumes where too small.

Finally, we note that a similar but independent study of $\calN=1$ SYM was presented by J. Giedt at this conference \cite{Giedt:2008aa}.

\section{Simulation and measurement details}

\begin{figure}
\begin{minipage}[t]{0.486\textwidth}
\centering
\includegraphics[angle=270,width=2.9in]{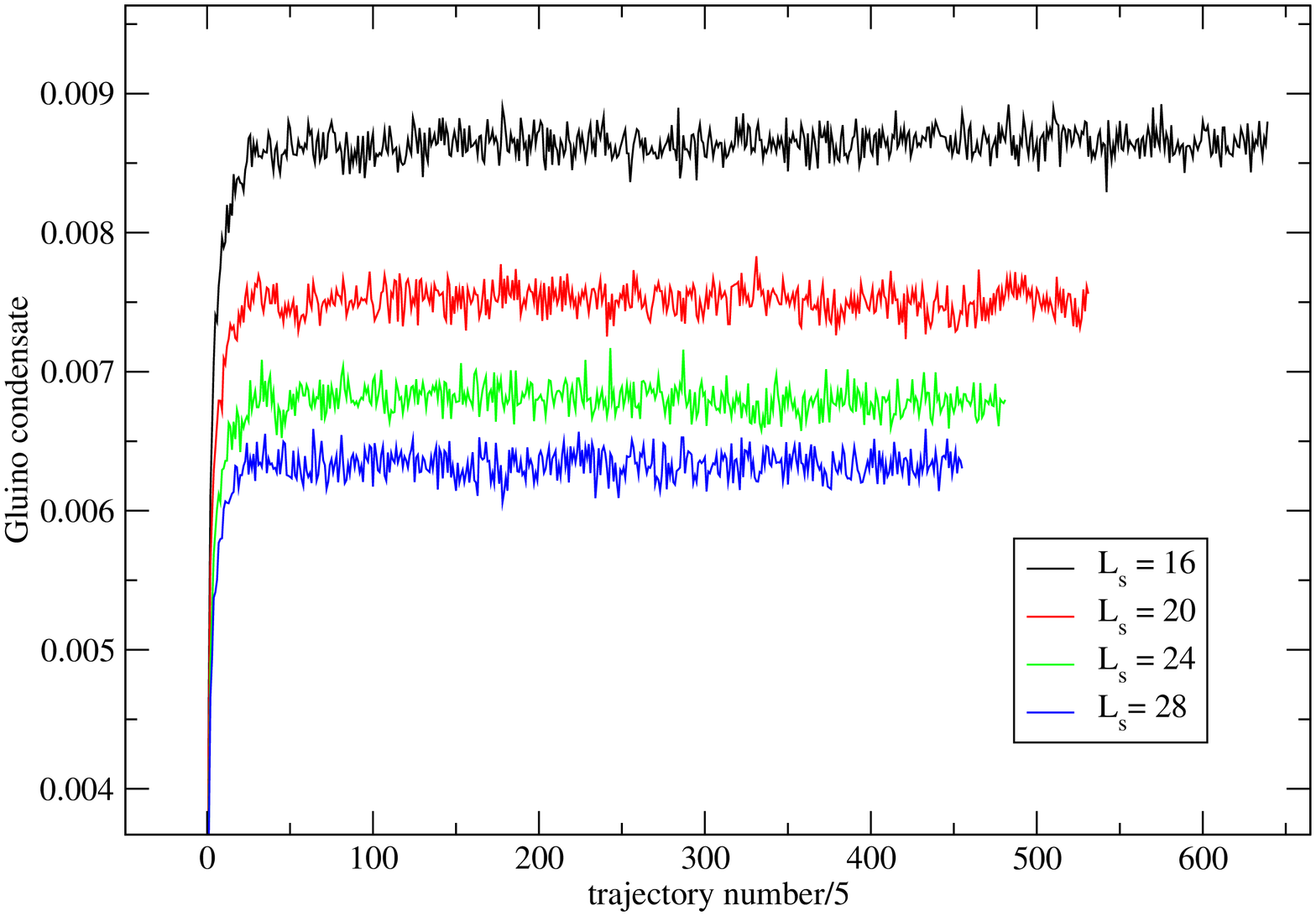}
\caption{Monte Carlo time history for the gluino condensate for $\beta=2.3$ and $m_f=0.02$.} 
\label{fig:thermalization}
\end{minipage}
\hspace{8pt}
\begin{minipage}[t]{0.486\textwidth}
\centering
\includegraphics[angle=270,width=2.9in]{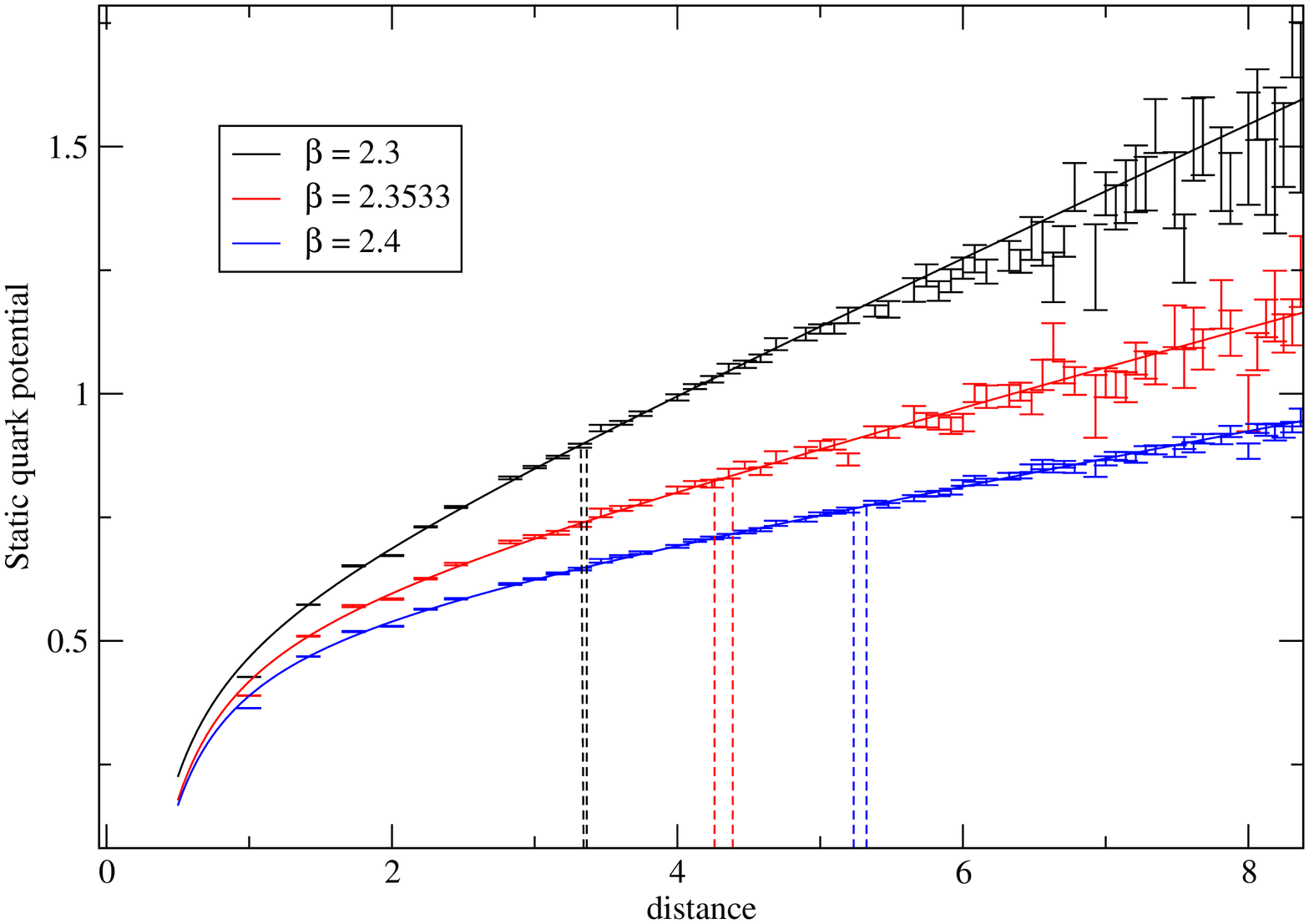}
\caption{%
Heavy quark potential as a function of distance for $m_f=0.02$ and $L_s=16$.
Sommer scale error bars are indicated by dashed lines.}
\label{fig:hqpot}
\end{minipage}
\end{figure}

Numerical simulations of $\calN=1$ SYM were performed using an appropriately modified version of the Columbia Physics System (CPS).
We use a Wilson gauge action with Majorana DWFs in the adjoint representation of the gauge group $SU(2)$.
%Numerical simulations of $\calN=1$ SYM were performed using a Wilson gauge action and Majorana DWFs in the adjoint representation of the gauge group $SU(2)$.
The specific details of this lattice action may be found in \cite{Fleming:2000fa}.
Rational Hybrid Monte Carlo simulations were performed on a $16^3\times32\times L_s$ lattice with $L_s=16,20,24$ and $28$, gluino masses $m_f = 0.01, 0.02$ and $0.04$, a domain-wall height of $M=1.9$, and coupling $\beta=2.3$.
Several additional simulations where performed at the weaker couplings $\beta = 2.3533$ and $\beta=2.4$ as well.
For each ensemble, a total of 2500 to 3000 trajectories were generated starting from an ``ordered'' configuration and equilibrium was achieved within the first $500$ trajectories.
Measurements were made using uncorrelated configurations which were generated thereafter.
A plot of the gluino condensate time history is shown in \Fig{thermalization}.

%%%%%%%%%%%%%%%%%%%%%%%%%%%%%%%%%%%%%%%%
%  Figure
%%%%%%%%%%%%%%%%%%%%%%%%%%%%%%%%%%%%%%%%
%%%%%%%%%%%%%%%%%%%%%%%%%%%%%%%%%%%%%%%%

\section{Results}

\subsection{Static quark potential}
The static quark potential was extracted from Wilson loop measurements for a range of couplings.
Wilson loops were measured in the fundamental representation of the gauge group and on Coulomb gauge fixed gauge configurations.
For fixed distances $r=|\bfx|$, the Wilson loops were fit as a function of time to the formula:
\beq
\langle W(\bfx,t) \rangle = C(\bfx) e^{-V(\bfx) t}\ , \qquad V(\bfx) = V_0 - \frac{\alpha}{|\bfx|} + \sigma |\bfx|
\eeq
within an interval were excited state contamination appeared to be negligible.
The extracted values of $V(\bfx)$ were then fit to the Cornell potential.
The constant term ($V_0$), L\"uscher term ($\alpha$), string tension ($\sigma$), and Sommer scale ($r_0$) defined by:
\beq
\left. |\bfx|^2 \frac{\partial V(\bfx)}{\partial |\bfx|} \right|_{|\bfx|=r_0} = 1.65
\eeq
were determined by double jackknife fits to the data.
\Tab{hqpot} and \Fig{hqpot} summarizes the fit results.
A decrease in the string tension with coupling supports the conclusion that we are in a confining phase of the theory.
Taking the Sommer scale to be $r_0 = 0.5$ fm, we find that the inverse lattice spacing ranges between 1.3 GeV at $\beta=2.3$ and 2.1 GeV at $\beta=2.4$.

\begin{table}[t]
\caption{Static quark potential fit parameters for $m_f=0.02$ and $L_s=16$.}
\centering
\begin{tabular}{l l c c c c c c c}
\hline\hline
$\beta$ & t range & r range  & $r_0$  & $V_0$ & $\alpha$ & $\sigma$ \\
\hline
2.3    & 4-8   & $\sqrt{3}$-6 & 3.339(23)  & 0.501(18) & 0.161(23) & 0.134(4) \\
2.3533 & 5-9   & $\sqrt{3}$-6 & 4.379(80)  & 0.569(23) & 0.240(23) & 0.074(4) \\
2.4    & 5-10  & $\sqrt{3}$-6 & 5.306(68)  & 0.539(11) & 0.205(15) & 0.051(2) \\
\end{tabular}
\label{tab:hqpot}
\end{table}

\subsection{Residual mass}

It is of crucial importance that we have an understanding of the residual mass ($m_{res}$) since its magnitude determines our proximity to the SUSY point in this lattice formulation.
The $L_s$ dependence of the residual mass may be parameterized by the theoretically motivated formula \cite{Antonio:2008zz}:
\beq
m_{res} \sim a_0 \frac{e^{- a_1 L_s}}{L_s} + a_2 \frac{\rho(0)}{L_s}\ ,
\label{eq:mres}
\eeq
where $\rho(0)$ appearing in the second term (i.e. the dislocation term) represents the density of zero eigenvalues of the fifth dimension transfer matrix Hamiltonian.

\begin{figure}
\begin{minipage}[t]{0.486\textwidth}
\centering
\includegraphics[angle=270,width=2.9in]{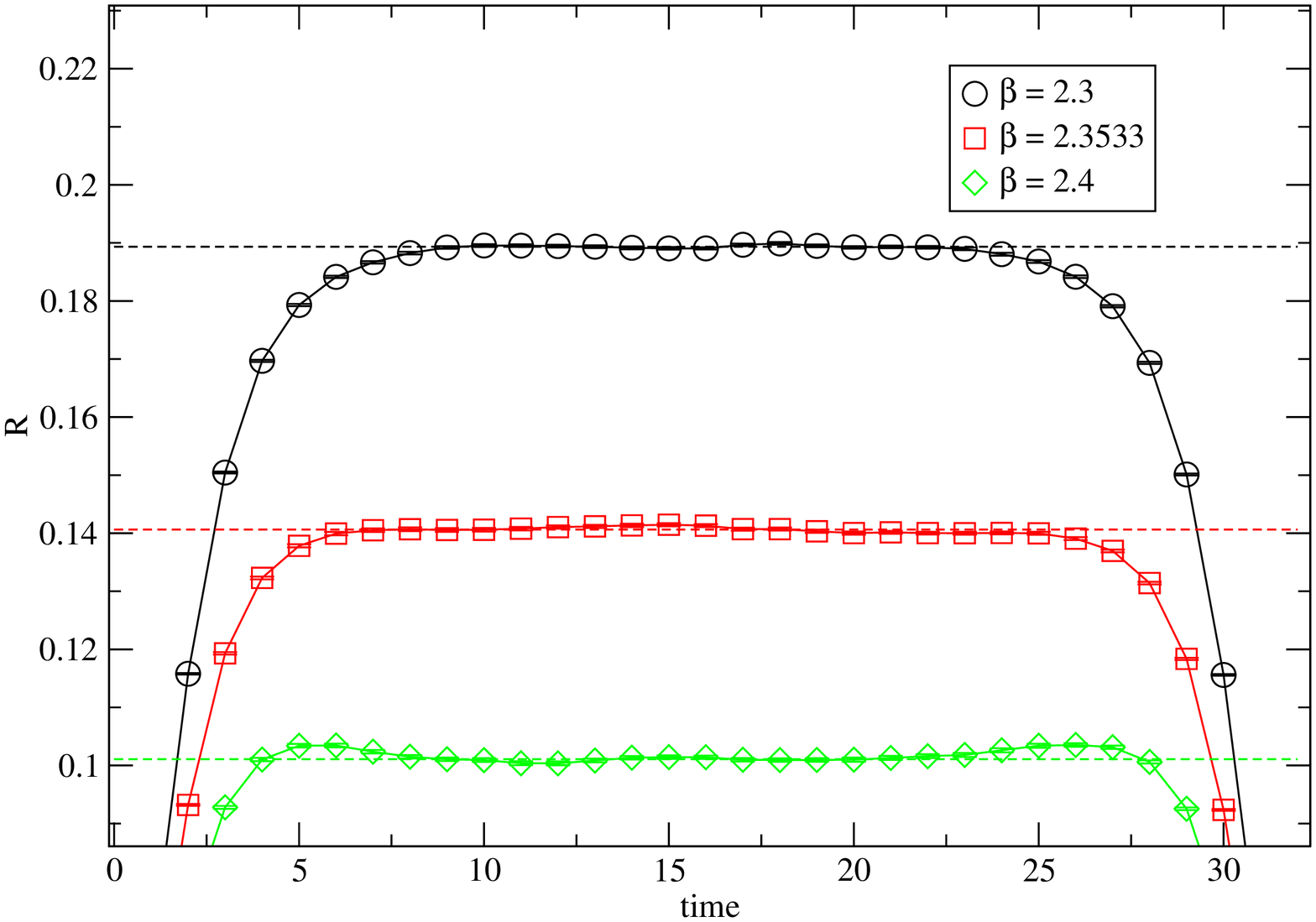}
\caption{R as a function of time for $m_f=0.02$ and $L_s=16$.} 
\label{fig:R}
\end{minipage}
\hspace{8pt}
\begin{minipage}[t]{0.486\textwidth}
\centering
\includegraphics[angle=270,width=2.9in]{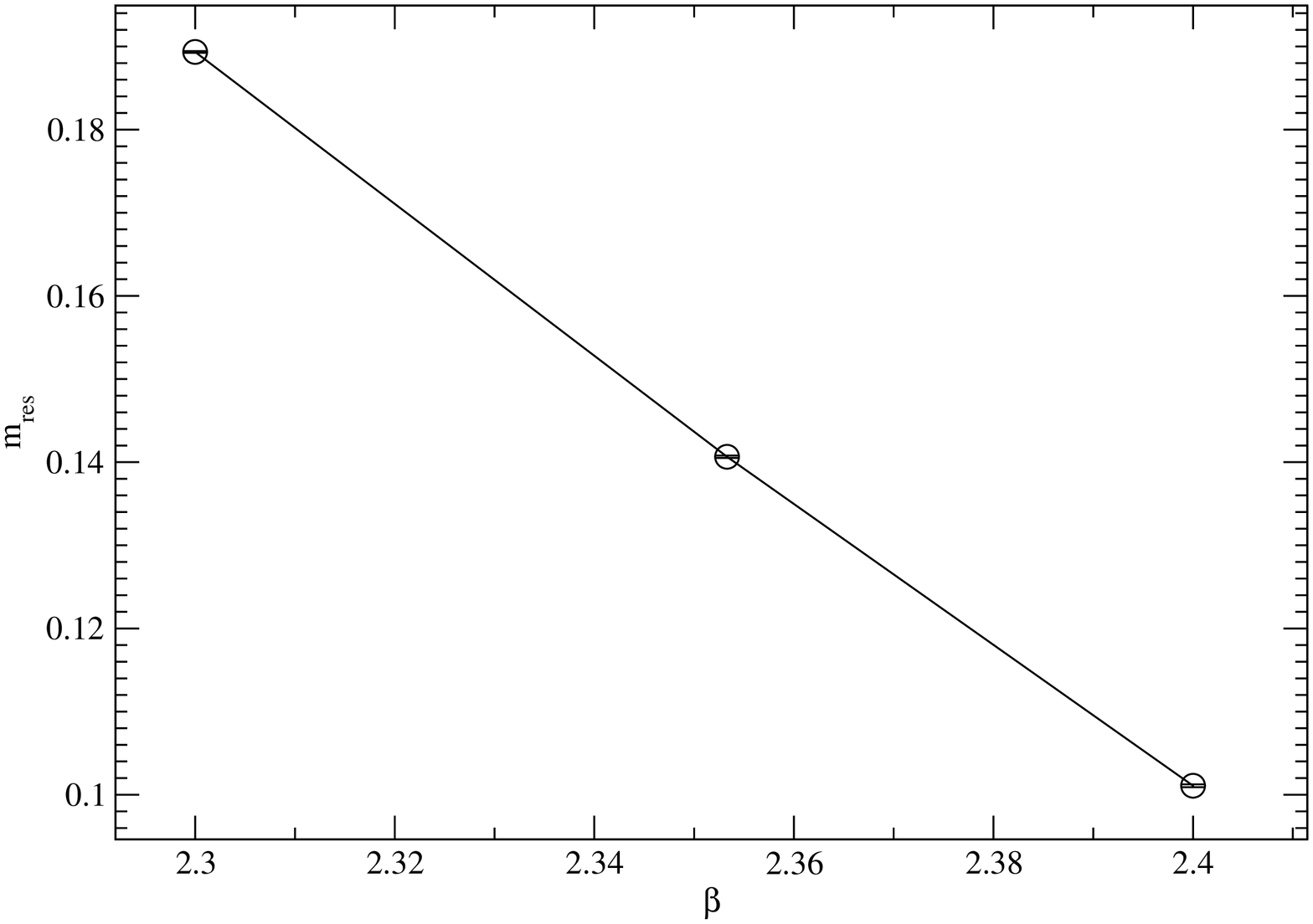}
\caption{Residual mass as a function of coupling for $m_f=0.02$ and $L_s=16$. }
\label{fig:mres}
\end{minipage}
\end{figure}

The residual mass was determined from a ratio $R(t)$, given by the coupling of the pion \footnote{Here, we refer to the flavor non-singlet pseudo-scalar associated with a partially quenched theory.} to the mid-point pseudo-scalar density divided by its coupling to the boundary (see \cite{Blum:2000kn} for details).
At large times this ratio of correlation functions is expected to tend asymptotically toward $m_{res}$; plots of this ratio for $L_s=16$ lattices are shown in \Fig{R}.
We determine the residual mass by fitting $R(t)$ with a constant over the plateau region.
\Fig{mres} shows a plot of the extracted values of $m_{res}$ as a function of the coupling.
We find that the residual mass is roughly 5-10 times that of the input gluino mass.
Furthermore, the strong coupling dependence of $m_{res}$ suggests that the dislocation term appearing in \Eq{mres} dominates the residual mass.
In order to reduce the residual mass, simulations at weaker couplings and larger $L_s$ values are currently underway.

%%%%%%%%%%%%%%%%%%%%%%%%%%%%%%%%%%%%%%%%
%  Figure
%%%%%%%%%%%%%%%%%%%%%%%%%%%%%%%%%%%%%%%%
%%%%%%%%%%%%%%%%%%%%%%%%%%%%%%%%%%%%%%%%

\subsection{Gluino condensate}
The gluino condensate was measured using a stochastic estimator with a single hit.
We perform chiral limit extrapolations of the gluino condensate using two different limit orders following \cite{Fleming:2000fa}.
First, we perform a linear $m_f \to0$ extrapolation of the gluino condensate at fixed $L_s$, followed by an $L_s \to \infty$ extrapolation of the $m_f=0$ result using the best available, theoretically motivated fit formula:
\beq
c_0 + c_1 \frac{e^{-c_2 L_s}}{L_s}\ ,
\label{eq:Ls_fit}
\eeq
which may be derived from the fifth dimension transfer matrix formalism.
Note that the dislocation contribution to $m_{res}$ which appears in \Eq{mres} is absent in \Eq{Ls_fit}.
This may be understood by observing that the chiral condensate is dominated by contributions from UV modes, whereas the dislocation term appearing in $m_{res}$ may be attributed to purely low energy phenomena \cite{Cheng:2008aa,RBC:2008aa}.

\begin{figure}
\begin{minipage}[t]{0.486\textwidth}
\centering
\includegraphics[angle=270,width=2.9in]{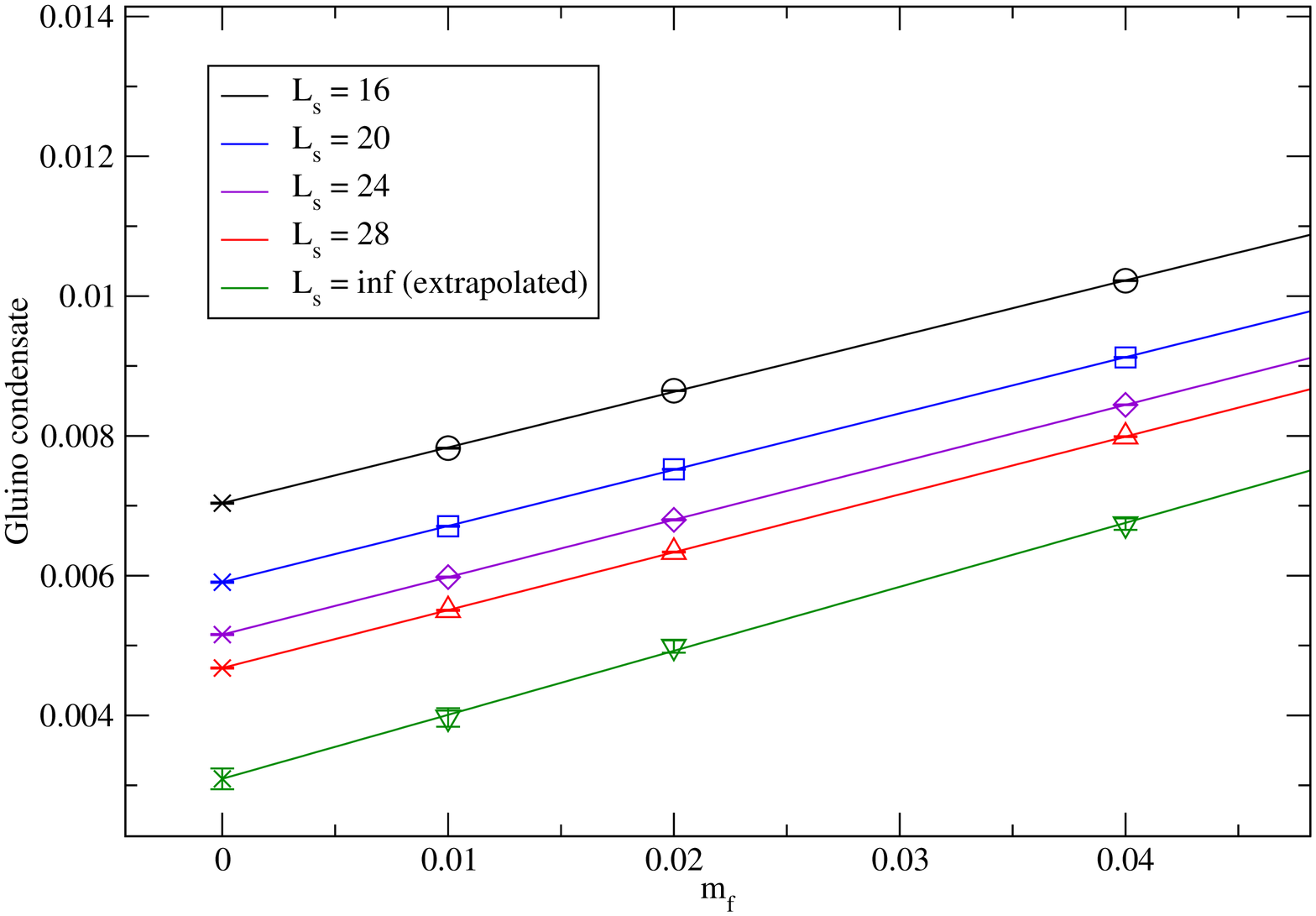}
\end{minipage}
\hspace{8pt}
\begin{minipage}[t]{0.486\textwidth}
\centering
\includegraphics[angle=270,width=2.9in]{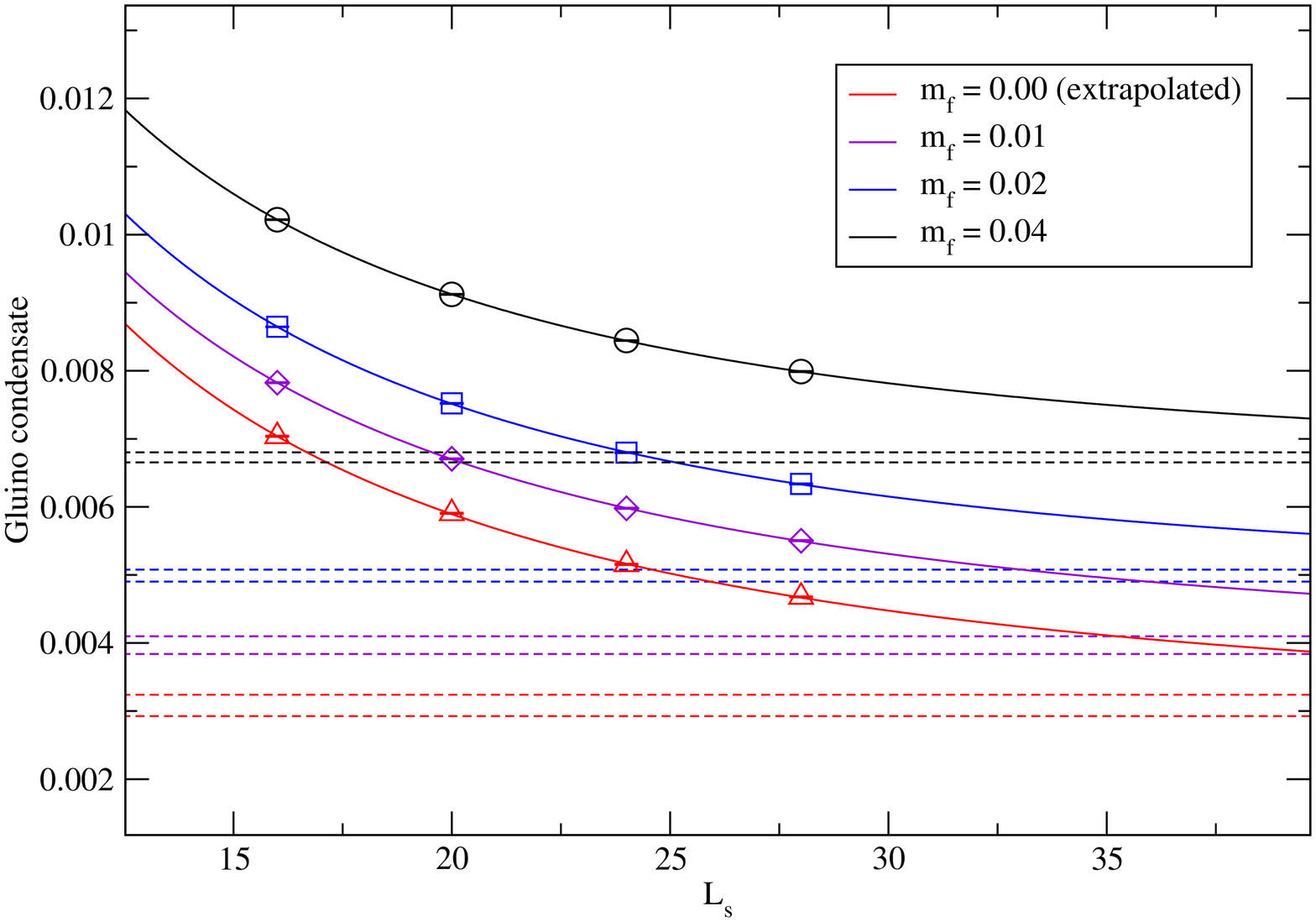}
\end{minipage}
\caption{Fits of the gluino condensate as a function of $m_f$ (left; extrapolated values are indicated by $\times$) and as a function of $L_s$ (right; extrapolation error bars are indicated by dashed lines) for $\beta = 2.3$.} 
\label{fig:condensate}
\end{figure}

Following the double extrapolation procedure outlined above we obtain an unrenormalized value of 0.003087(159) ($\chi^2/d.o.f. = 3.3$) for the gluino condensate in the chiral limit at finite lattice spacing.
Chiral extrapolations have been performed by reversing the order of limits and yield consistent results with comparable error bars.
Plots of the $m_f$ and $L_s$ fits for these extrapolations are shown in \Fig{condensate}.
We have performed additional fits using other, phenomenologically motivated fit formulae to describe the $L_s$ dependence of the gluino condensate (e.g. \Eq{Ls_fit}, without the $L_s^{-1}$ prefactor in the second term).
Such fits yield a 20\% shift in the gluino condensate as compared to the value obtained with \Eq{Ls_fit}.

%%%%%%%%%%%%%%%%%%%%%%%%%%%%%%%%%%%%%%%%
%  Figure
%%%%%%%%%%%%%%%%%%%%%%%%%%%%%%%%%%%%%%%%
%%%%%%%%%%%%%%%%%%%%%%%%%%%%%%%%%%%%%%%%

\subsection{Spectrum}

The low energy spectrum of $\calN=1$ SYM is believed to consist of supermultiplets which may involve glue-glue, glue-gluino as well as gluino-gluino bound states.
Although the states within a given multiplet are degenerate, at finite gluino mass (e.g. $L_s\neq\infty, m_f\neq0)$ one expects mass splittings which are to leading order linear in the gluino mass.
The mass splittings have been calculated using a variety of effective theories \cite{Veneziano:1982ah,Farrar:1997fn}, however such calculations are unreliable since there is no separation of scales and therefore no small expansion parameter. 

For the scalar and pseudo-scalar gluino-gluino and the glue-gluino composite states, we consider the interpolating fields
\beq
\Omega_i(\bfx,t) = \Tr \bar \lambda(\bfx,t) \Gamma_i \lambda(\bfx,t)\ , \qquad \Omega(\bfx,t) = \Tr F_{\mu\nu}(\bfx,t) \Sigma_{\mu\nu} \lambda(\bfx,t)\ , 
\eeq
where $\Gamma_i = \{1, \gamma_5 \}$ for the scalar (s) and pseudo-scalar (ps), $\lambda(\bfx,t)$ is an appropriate interpolating field for the gluino, $\Sigma_{\mu\nu} = \frac{-i}{2} [ \gamma_\mu, \gamma_\nu ]$ and $F_{\mu\nu}$ represents an interpolating field for the field strength tensor (e.g. a clover-leaf shaped product of link matrices).
The lowest energy states created by the former operator correspond to the $f_0$ and $\eta^\prime$ respectively in QCD, whereas there is no QCD analogue for the latter operator.
Upon ``integrating out'' the fermion degrees of freedom, the correlation function for the scalar and pseudo-scalar operators involve a difference between two contributions: a ``connected'' part $C_i(t)$ and a ``disconnected'' part $D_i(t)$.
As is the case with the $\eta^\prime$ in QCD, the disconnected contribution is numerically extremely difficult to evaluate exactly.
We choose to instead use a stochastic estimator to approximate the correlator following the techniques of \cite{Hashimoto:2008xg}.

\begin{table}[t]
\caption{Pseudo-scalar fit parameters for $m_f=0.02$ and $L_s=16$.}
\centering
\begin{tabular}{l|cc|cc}
\hline\hline
$\beta$ & t range & $m_{ps}^{connected}$ & t range & $\Delta m_{ps}$ \\
\hline
2.3    & 9-23 & 0.8701(3) & 2-4 & 0.0180(37) \\
2.3533 & 9-23 & 0.8144(6) & 2-4 & 0.0176(52) \\
2.4    & 9-23 & 0.7367(9) & 2-4 & 0.0230(51) \\
\end{tabular}
\label{tab:spectrum}
\end{table}

\begin{figure}
\begin{minipage}[t]{0.486\textwidth}
\centering
\includegraphics[angle=270,width=2.9in]{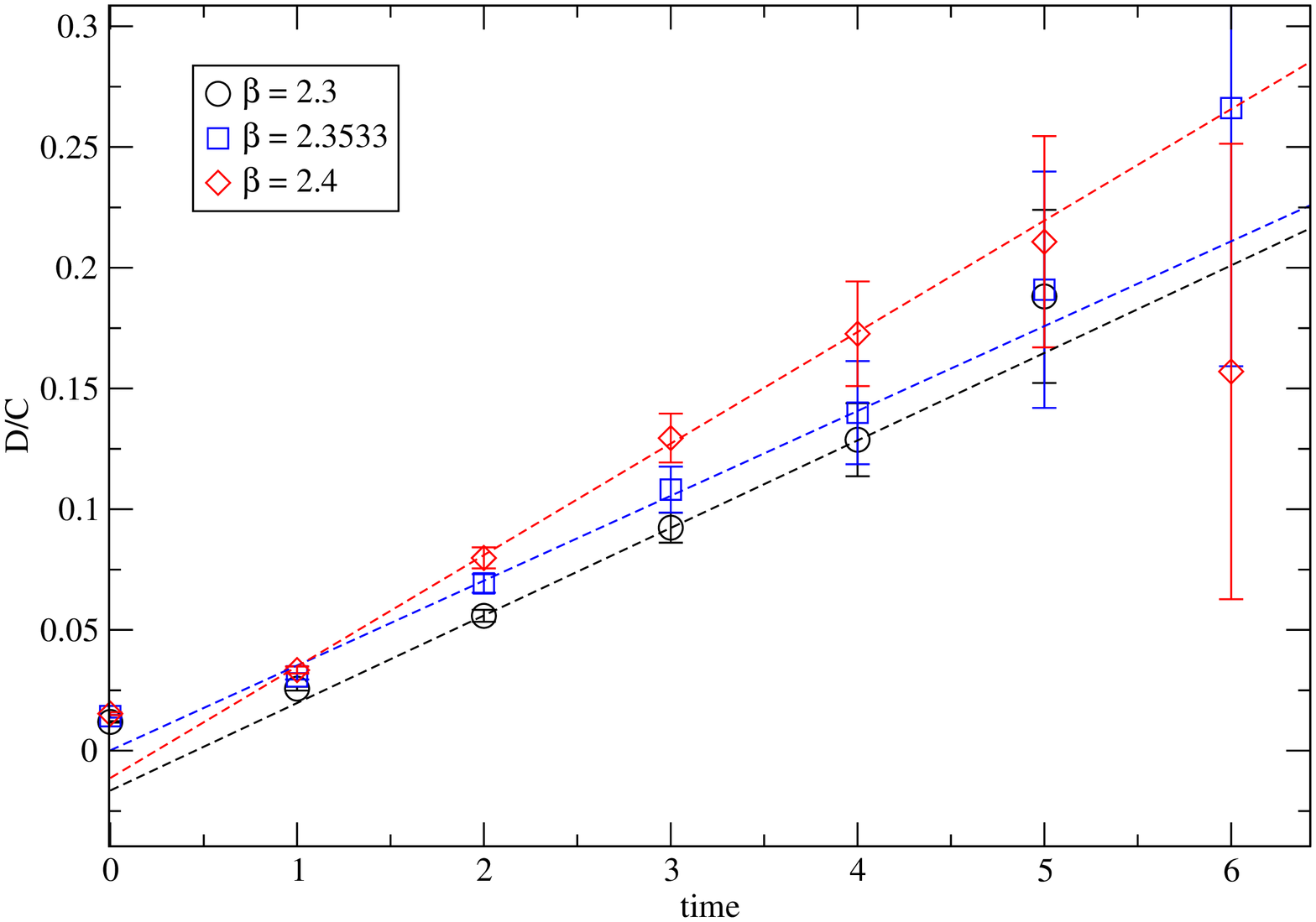}
\caption{Ratio of connected and disconnected pseudo-scalar correlators as a function of time for $m_f = 0.02$ and $L_s = 16$.} 
\label{fig:ratio}
\end{minipage}
\hspace{8pt}
\begin{minipage}[t]{0.486\textwidth}
\centering
\includegraphics[angle=270,width=2.9in]{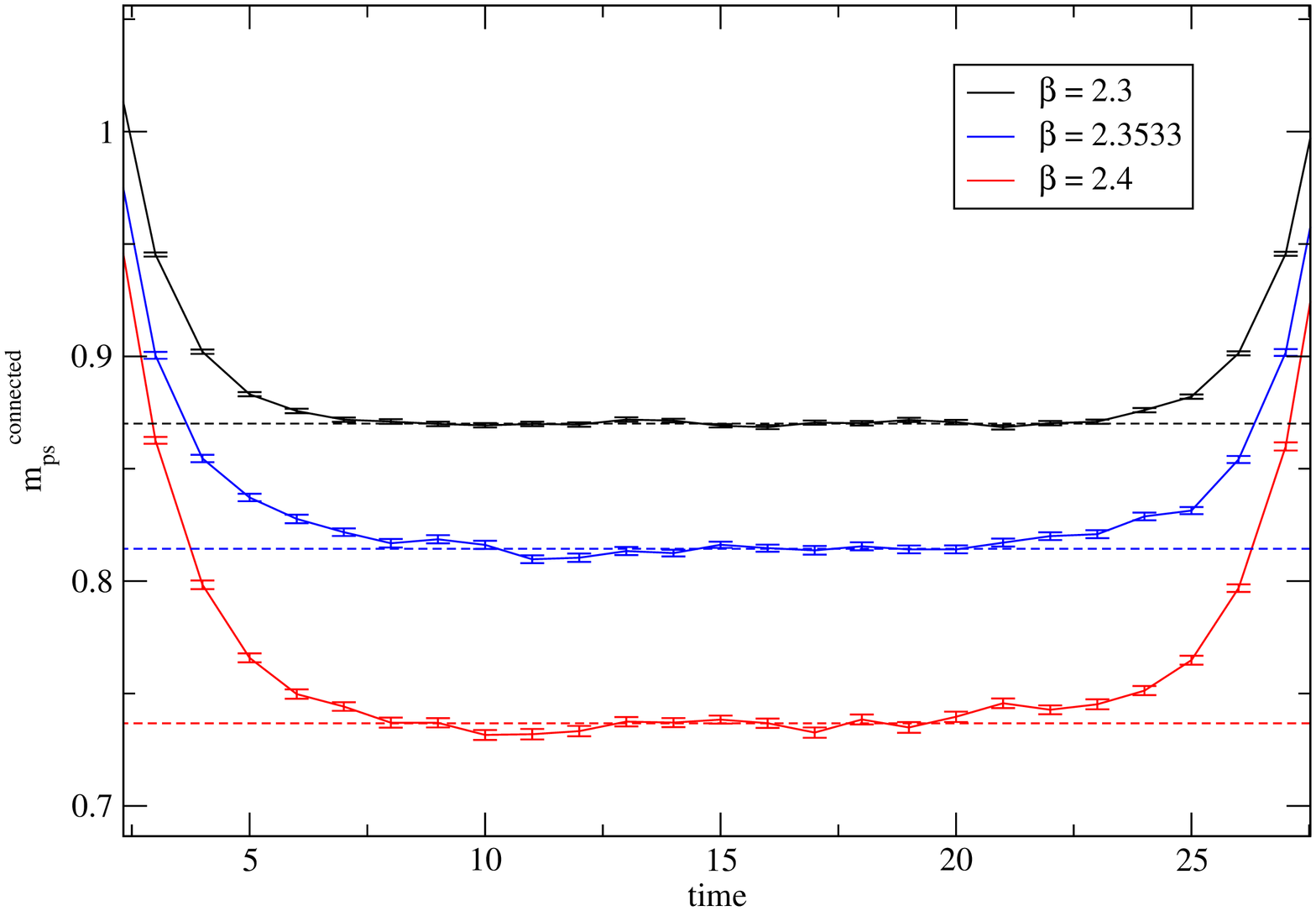}
\caption{Effective mass plot of the connected pseudo-scalar correlator as a function of time for $m_f = 0.02$ and $L_s = 16$.} 
\label{fig:meff}
\end{minipage}
\end{figure}

The connected and disconnected correlators where measured using random $Z_2$ volume and wall sources respectively (a single hit for the connected part and five hits for the disconnected part); to improve statistics we project onto the zero momentum state by averaging the result over all of space.
In order to extract the mass of the pseudo-scalar, we study the ratio of disconnected and connected correlation functions 
\beq
\frac{D_{ps}(t) }{C_{ps}(t) } = 2 - d e^{-\Delta m_{ps} t}\ ,
\label{eq:ratio}
\eeq
where $\Delta m_{ps} = m_{ps} - m_{ps}^{connected}$.
Plots of this ratio at several different couplings are shown in \Fig{ratio}.
The linearity of these plots appear consistent with $\Delta m_{ps} t \ll 1$, presumably due to the presence of a large residual mass.
Assuming that this is the case, we may expand \Eq{ratio} to leading order in $\Delta m_{ps} t$ and then perform a linear fit to obtain a value for the mass difference $\Delta m_{ps}$.
\Fig{meff} shows effective mass plots for the connected part of the pseudo-scalar correlation function at several different couplings from which $m_{ps}^{connected}$ may be extracted.
With $\Delta m_{ps}$ and $m_{ps}^{connected}$ determined, we may finally extract the pseudo-scalar mass $m_{ps}$.
The results of these fits are provided in \Tab{spectrum}.
While at present there are insufficient statistics to differentiate $\Delta m_{ps}$ between $\beta$ runs, it is none-the-less evident that for each coupling $m_{ps}$ is dominated by its contribution from $m_{ps}^{connected}$.
A complete analysis of the pseudo-scalar, scalar and their fermionic superpartner at smaller residual masses is currently underway and results will appear in a forthcoming publication.

%%%%%%%%%%%%%%%%%%%%%%%%%%%%%%%%%%%%%%%%
%  Figure
%%%%%%%%%%%%%%%%%%%%%%%%%%%%%%%%%%%%%%%%
%%%%%%%%%%%%%%%%%%%%%%%%%%%%%%%%%%%%%%%%

\section{Acknowledgments}
M. G. E. would like to thank N. Christ, C. Kim and R. Mawhinney for numerous helpful discussions, I. Mihailescu for fitting the static quark potential data presented in this work, and C. Jung for technical assistance with compiling and running CPS on BlueGene/L.
This research utilized resources at the New York Center for Computational Sciences at Stony Brook University/Brookhaven National Laboratory which is supported by the U.S. Department of Energy under Contract No. DE-AC02-98CH10886 and by the State of New York.
This work was supported by the U.S. Department of Energy under grant number DE-FG02-92ER40699.

\bibliography{lattice2008}
\bibliographystyle{h-physrev}

\end{document}